\documentclass[aps,preprint]{revtex4}

\usepackage{graphics,graphicx}
\usepackage{amsmath,amssymb}
\usepackage{epsfig}
\usepackage{txfonts}
\usepackage{subfigure}
\usepackage{xcolor}
\usepackage{epstopdf}
\usepackage{hyperref}
\begin{document}

\title{$q$-form field on a $p$-brane with codimension two}
\author{Zi-Qi Chen$^{1}$\footnote{Email: chenzq16@lzu.edu.cn}}
\author{Chun-E Fu $^{2,}$\footnote{Email: fuche13@mail.xjtu.edu.cn}}
\author{Chen Yang $^{1}$\footnote{Email: yangch2017@lzu.edu.cn}}
\author{Li Zhao $^{1}$\footnote{Email: lizhao@lzu.edu.cn, corresponding author}}

\affiliation{\small{$^{1}$ Institute of Theoretical Physics, Lanzhou University, 222 South Tianshui Road, Lanzhou 730000, China\\
                    $^{2}$ School of Science, Xi'an Jiaotong University, Xi'an 710049, China\\
                  }}

\begin{abstract}
In this paper,  by performing a general Kaluza-Klein
(KK) decomposition, we obtain a gauge invariant effective action for a bulk massless
$q$-form field on a $p$-brane with codimension two. There appear four types of KK modes:
two $(q-1)$-forms and one $(q-2)$-form in addtion to the ordinary $q$-form, which are essential for the gauge invariance.
Due to the two extra dimensions, we find eight Schr\"{o}dinger-like equations for the four modes and their mass spectra are closely related.
Moreover, via this decomposition mechanism, the Hodge duality in the bulk naturally induces four coupled dualities on the brane, which guarantees that the physical equavalence of bulk dual fields is preserved under the dimensional reduction.
\end{abstract}

%\pacs{04.50.Kd, 98.80.-k}

\maketitle

\section{Introduction}
In the extra-dimensional theories, the Kaluza-Klein (KK) theory provides a possibility of  unification of electromagnetism and gravity by introducing one compact extra dimension. This  idea has attracted new attention and been widely studied \cite{Haghani2012,George2009,DeWolfe2000,Gregory2000,Kaloper2000,Wang2002,Kobayashi2002,Melfo2003,Cardoso2007,Dzhunushaliev2008,
Bazeia2004,Bazeia2006,Liu2010,Liu2011,Liu2012,Liu2013,Bazeia2014,Das2014} since the Arkani-Hamed-Dimopoulos-Dvali (ADD) model \cite{Arkani-Hamed1998} and the Randall-Sundrum (RS) model \cite{Randall1999,Sundrum1999} provided new ways to solve the long-existing hierarchy problem \cite{Randall1999,Yang2012,Antoniadis1998,Arkani-Hamed1998,Gogberashvili2002} and the cosmology problem \cite{Arkani-Hamed2000,Neupane2011,Starkman2001,Kim2001}.

In the braneworld scenario, the KK modes of bulk fields are of particular importance \cite{Grossman2000,Bajc2000,Gremm2000,Chang2000,Randjbar-Daemi2000,Kehagias2001,Duff2001,Oda2001,Ringeval2002,Ichinose2002,Koley2005,Davies2008,Liu2008,
Archer2011,Jones2013,Xie2013,Cembranos2013,Costa2013,Zhao2014,Kulaxizi2014}. Through a reduction mechanism, various bulk fields naturally generate a series of KK modes. The zero modes should coincide with particles in our observed 4D spacetime, while the massive KK modes reveal the physics of the extra dimensions. In this work we consider a massless $q$-form field in a $D=p+3$ dimensional spacetime bulk where a codimensional two brane resides, i.e., the brane has p spatial dimensions. The 0-form and 1-form fields are respectively the well-known scalar and vector fields. A 2-form field is the Kalb-Ramond field, which appears as the torsion of the spacetime in the Einstein-Cartan theory and as a massless mode in the string theory as well. Higher $q$-form fields ($q>2$) only exist in high dimensional spacetime with $D>q+1$ and represent new particles with $D>q+4$ , so as to avoid a vacuous discussion we do not restrict our $p$-brane to $p=3$. This is plausible for, in principle, there can be compact small-scale dimensions besides the ordinary 3 infinite dimensions, and together they form the $p$-brane.

There has been some work on localization of the $q$-form fields \cite{Duff2001,Rindani1985,Sivakumar1988,Mukhopadhyaya2002,Mukhopadhyaya,Alencar2010,Youm2001,Fu2012,Fu2014,Ho2013,Hinterbichler2014,
Smailagic2000}. In order to obtain a series of KK modes of the bulk field, one carries out a dimensional reduction with some localization mechanism. Under the usual mechanism, the KK decomposition is performed after a gauge fixing to simplify the derivation. This eventually causes a problem: it is known that a $q$-form field and its Hodge dual, a ($D-2-q$)-form field, are physically equivalent, while via this mechanism, only one of them is localizable. In order to eliminate this unreasonable contradiction,
modifications have been proposed in some work, one of which is of particular interest: the authors
of \cite{Fu2016} suggested a general KK decompodition without gauge-fixing, under which the Hodge duality in the bulk naturally transfers onto the brane. With this new idea, another work discussed the localization of a vector field on a codimension-two brane \cite{Fu2019}.  Here we will follow them to study the more general case of $q$-form on a $p$-brane of codimension-two and reconsider the duality issue, and this strongly indicates the results do not depend on the number of the codimensions either.

In more detail, we start by carrying the general KK decomposition to a bulk $q$-form $X$, which gives four types of KK modes $X^{(n)}$, $\overline{X}^{(n)}$, $\underline{X}^{(n)}$, $\overline{\underline{X}}^{(n)}$. One of them is the usual $q$-form mode $X^{(n)}$, the other three are forms of lower rank ($q-1$, $q-1$, $q-2$ respectively), which typically exist in this new mechanism and are essential for the  gauge-invariance of the brane action and brane Hodge duality. By assuming orthogonality conditions and employing a technique of comparing two sets of equations, we can seperate equations of the KK modes into brane parts and extra parts. For each type of extra dimensional functions, there are two Schr\"{o}dinger-like equations that correspond to the two codimensions, thus each type gets two parts of masses from extra dimensions. In total we have eight Schr\"{o}dinger-like equations, and their mass spectra are closely related as the KK modes are coupling with each other. In contrast to the codimension-one case, these are partial differential equations, and we do not analysis their solutions since our discussion is formalism.

As usual, in the brane action obtained from KK reduction, three of the fields have mass terms, which are considered as obstacles to gauge-invariance of the action. Fortunately,  we are able to reformulate the brane action and find out that the four brane fields couple in a nice way so that under proper gauge transformations they compensate each other, which eventually make the action invariant. Despite its extra dimensional origin, this is similar to the Higgs mechanism where via coupling with a scalar, a massless vector gains mass without loss of gauge invariance.

By virtue of this new mechanism, Hodge duality is also preserved under the codimension-two reduction. We will find two dual forms in the bulk of ranks $q$ and $p+1-q$, i.e, of strength ranks $q+1$ and $p+2-q$, naturally lead to four pairs of coupled dualities between brane form fields, of ranks  $q-1 \sim p+2-q$, $q \sim p+1-q$, $q \sim p+1-q$ and $q+1 \sim p-q$, respectively. They are exactly the counterparts appear in the effective brane action, therefore the brane action equals to its dual action. In this sense the equivalent bulk fields reduce to equivalent brane fields. Moreover, the previous localizability contradiction will automatically disappear due to some relations between extra-dimensional functions.

This paper is organized as follows. We perform a general KK decomposition and obtain four types of KK modes in Sec.\ref{sec:model}, and then discuss the gauge invariance of the brane action in Sec.
\ref{invariance}. In Sec. \ref{Hodge dualities} we show the  Hodge duality in the
bulk and on the brane. Finally, we give a conclusion and discussion in Sec. \ref{conclusion}.

\section{A general Kaluza-Klein decomposition}
\label{sec:model}
We specify the metric $ds^{2}=e^{2A(y,z)}\left(\hat{g}_{\mu\nu}(x^{\lambda})dx^{\mu}dx^{\nu}+dy^{2}+dz^{2}\right)$ for our discussion, where $\hat{g}_{\mu\nu}$ is the induced metric on the (p+1)-dimensional brane world, and $A(y,z)$ is the warp factor which only depends on the two extra dimensional coordinates.
Consider a massless $q$-form field  $X_{M_{1} \cdots M_{q}}$ in the bulk, and the action reads
\begin{eqnarray}
\mathrm{S}&=&-\frac{1}{2(q+1) !} \int d^{D} x \sqrt{-g} Y^{M_{1} \cdots M_{q+1}} Y_{M_{1} \cdots M_{q+1}},\nonumber \\
&=&-\frac{1}{2(q+1) !} \int d^{D} x \sqrt{-g} \left[Y^{\mu_{1} \cdots \mu_{q+1}} Y_{\mu_{1} \cdots \mu_{q+1}}+(q+1) Y^{\mu_{1} \cdots \mu_{q} y} Y_{\mu_{1} \cdots \mu_{q} y}\right. \nonumber \\
&&\left.+(q+1) Y^{\mu_{1} \cdots \mu_{q} z} Y_{\mu_{1} \cdots \mu_{q} z}+q(q+1) Y^{\mu_{1} \cdots \mu_{q-1} y z} Y_{\mu_{1} \cdots \mu_{q-1} y z}\right],\label{action}
\end{eqnarray}
where $Y_{M_{1} \cdots M_{q+1}}=\partial_{[M_{1}}X_{M_{2} \cdots M_{q+1}]}$   is the field strength of $X_{M_{1} \cdots M_{q}}$.
The equations of motion (EoMs) for the bulk field $\partial_{M_{1}}\left(\sqrt{-g} Y^{M_{1} \cdots M_{q+1}}\right)=0$ are then
\begin{eqnarray}&&{\partial_{\mu_{1}}\left(\sqrt{-g} Y^{\mu_{1} \cdots \mu_{q-1}y z}\right)=0}, \nonumber\\
 &&{\partial_{\mu_{1}}\left(\sqrt{-g} Y^{\mu_{1} \cdots \mu_{q} z}\right)+\partial_{y}\left(\sqrt{-g} Y^{y \mu_{2} \cdots \mu_{q} z}\right)=0},\nonumber \\ &&{\partial_{\mu_{1}}\left(\sqrt{-g} Y^{\mu_{1} \cdots \mu_{q} y}\right)+\partial_{z}\left(\sqrt{-g} Y^{z \mu_{2} \cdots \mu_{q} y}\right)=0},\nonumber \\ &&{\partial_{\mu_{1}}\left(\sqrt{-g} Y^{\mu_{1} \cdots \mu_{q+1}}\right)+\partial_{y}\left(\sqrt{-g} Y^{y \mu_{2} \cdots \mu_{q+1}}\right)+\partial_{z}\left(\sqrt{-g} Y^{z \mu_{2} \cdots \mu_{q+1}}\right)=0}.
\end{eqnarray}
Instead of using the usual  KK decomposition with the specific gauge condition, we follow the method developed in Ref. \cite{Fu2016}, where a general decomposition is assumed to preserve the full information of the extra dimension. We start with the general gauge-free KK decomposition for the bulk $q$-form field
\begin{eqnarray}
{\mathrm{X}_{\mu_{1} \cdots \mu_{q}}\left(x_{\mu}, y, z\right)=\sum_{n} X_{\mu_{1} \cdots \mu_{q}}^{(n)}\left(x^{\mu}\right) W_{1}^{(n)}(y, z) e^{a A(y, z)}}, \nonumber \\
 {\mathrm{X}_{\mu_{1} \cdots \mu_{q-1} y}\left(x_{\mu}, y, z\right)=\sum_{n} \overline{X}_{\mu_{1} \cdots \mu_{q-1}}^{(n)}\left(x^{\mu}\right) W_{2}^{(n)}(y, z) e^{a A(y, z)}}, \nonumber\\
  {{X}_{\mu_{1} \cdots \mu_{q-1} z}\left(x_{\mu}, y, z\right)=\sum_{n} \underline{X}_{\mu_{1} \cdots \mu_{q-1}}^{(n)}\left(x^{\mu}\right) W_{3}^{(n)}(y, z) e^{a A(y, z)}}, \nonumber\\
   {\mathrm{X}_{\mu_{1} \cdots \mu_{q-2} y z}\left(x_{\mu}, y, z\right)=\sum_{n} \overline{\underline{X}}_{\mu_{1} \cdots \mu_{q-2}}^{(n)}\left(x^{\mu}\right) W_{4}^{(n)}(y, z) e^{a A(y, z)}},
\end{eqnarray}
where $W_i^{n}$s \;($i=1,2,3,4$) are functions related to the extra dimensional coordiantes. Thus we have decomposed the bulk field potential $X$ into four types of KK mode fields $X^{(n)}$, $\overline{X}^{(n)}$, $\underline{X}^{(n)}$, $\overline{\underline{X}}^{(n)}$  on the brane,  which are respectively $q,q-1,q-1,q-2$ forms. Together with their field strengths $Y^{(n)}$, $\overline{Y}^{(n)}$, $\underline{Y}^{(n)}$, $\overline{\underline{Y}}^{(n)}$, their indices are raised or lowered by the induced  brane metric $\hat{g}_{\mu\nu}$. Here  $a$ is an arbitrary parameter, we choose it to be $q-(p+1)/2$ for later convenience. The KK decomposition of $X$ yields the decomposition of field strength $Y$ as
\begin{eqnarray}
 Y^{\mu_{1} \cdots \mu_{q+1} }&=&\sum_{n} Y_{(n)}^{\mu_{1} \cdots \mu_{q+1}} W_{1}^{(n)} e^{(a-2 q-2) A},\nonumber \\
 Y^{\mu_{1} \cdots \mu_{q} y}&=&\sum_{n}\left[\frac{(-1)^{q} e^{-2(q+1) A}}{q+1} X_{(n)}^{\mu_{1} \cdots \mu_{q}} \partial_{y}\left(W_{1}^{(n)} e^{a A}\right)+\frac{q e^{-2(q+1) A}}{q+1} \overline{Y}_{(n)}^{\mu_{1} \cdots \mu_{q}} W_{2}^{(n)} e^{a A}\right],\nonumber\\
Y^{\mu_{1} \cdots \mu_{q} z}&=&\sum_{n}\left[\frac{(-1)^{q} e^{-2(q+1) A}}{q+1} X_{(n)}^{\mu_{1} \cdots \mu_{q}} \partial_{z}\left(W_{1}^{(n)} e^{a A}\right)+\frac{q e^{-2(q+1) A}}{q+1} \underline{Y}_{(n)}^{\mu_{1} \cdots \mu_{q}} W_{3}^{(n)} e^{a A}\right],\nonumber\\
Y^{\mu_{1} \cdots \mu_{q-1} y z}&=&\sum_{n}\left[\frac{(-1)^{q-1} e^{-2(q+1) A}}{q+1} {{\underline{X}}}_{(n)}^{\mu_{1} \cdots \mu_{q-1}} \partial_{y}\left(W_{3}^{(n)} e^{a A}\right)\right.\nonumber\\
&&\left.+\frac{(-1)^{q} e^{-2(q+1) A}}{q+1} {\overline{X}}_{(n)}^{\mu_{1} \cdots \mu_{q-1}} \partial_{z}\left(W_{2}^{(n)} e^{a A}\right)+\frac{(q-1) e^{-2(q+1) A}}{(q+1)} {\overline{\underline{Y}}}_{(n)}^{\mu_{1} \cdots \mu_{q-1}} W_{4}^{(n)} e^{a A}\right]. \label{field strength1}
\end{eqnarray}
In order to obtain the $(p+1)$-dimensional effective brane action, we plug the decomposition of field strength $Y$ into Eq. (\ref{action}),
and the action reads,
\begin{eqnarray}
S=&&-\frac{1}{2(q+1) !} \sum_{n} \sum_{n^{\prime}} \int d^{p+1}x \sqrt{-\hat{g}}\quad
\left[I_{1}^{n n^{\prime}} Y_{(n)}^{\mu_{1} \cdots \mu_{q+1}} Y_{\mu_{1} \cdots \mu_{q+1}}^{\left(n^{\prime}\right)}+\left(I_{2}^{n n^{\prime}}+I_{4}^{n n^{\prime}}\right)
X^{n_{1} \cdots \mu_{q}}_{(n)} X_{\mu_{1}^{\prime} \cdots \mu_{q}}^{\left(n^{\prime}\right)}
\right.\nonumber\\
&& +I_{3}^{nn^{\prime}} \overline{Y}_{(n)}^{\mu_{1} \cdots \mu_{q}} \overline{Y}_{\mu_{1} \cdots \mu_{q}}^{\left(n^{\prime}\right)}
+I_{5}^{nn^{\prime}} \underline{Y}_{(n)}^{\mu_{1} \cdots \mu_{q}} \underline{Y}_{\mu_{1} \cdots \mu_{q}}^{\left(n^{\prime}\right)}
+2 I_{6}^{n n^{\prime}} X_{(n)}^{\mu_{1} \cdots \mu_{q}} \overline{Y}_{\mu_{1} \cdots \mu_{q}}^{\left(n^{\prime}\right)}
+2 I_{8}^{n n^{\prime}} X_{(n)}^{\mu_{1} \cdots \mu_{q}} \underline{Y}_{\mu_{1} \cdots \mu_{q}}^{\left(n^{\prime}\right)}\nonumber\\
&&+I_{7}^{n n^{\prime}} \overline{X}_{(n)}^{\mu_{1} \cdots \mu_{q-1}} \overline{X}_{\mu_{1} \cdots \mu_{q-1}}^{\left(n^{\prime}\right)}+I_{9}^{n n^{\prime}} \underline{X}_{(n)}^{\mu_{1} \cdots \mu_{q-1}} \underline{X}_{\mu_{1} \cdots \mu_{q-1}}^{\left(n^{\prime}\right)}
+2I_{10}^{n n^{\prime}} \overline{X}_{(n)}^{\mu_{1} \cdots \mu_{q-1}} \underline{X}_{\mu_{1} \cdots \mu_{q-1}}^{\left(n^{\prime}\right)}
+I_{11}^{n n^{\prime}} \overline{\underline{Y}}_{(n)}^{\mu_{1} \cdots \mu_{q-1}} \overline{\underline{Y}}_{\mu_{1} \cdots \mu_{q-1}}^{\left(n^{\prime}\right)}\nonumber\\
&&\left.+2I_{12}^{n n^{\prime}} \overline{X}_{(n)}^{\mu_{1} \cdots \mu_{q-1}} \overline{\underline{Y}}_{\mu_{1} \cdots \mu_{q-1}}^{\left(n^{\prime}\right)}
+2I_{13}^{n n^{\prime}} \underline{X}_{(n)}^{\mu_{1} \cdots \mu_{q-1}} \overline{\underline{Y}}_{\mu_{1} \cdots \mu_{q-1}}^{\left(n^{\prime}\right)}
 \right]\label{effectiveaction},
\end{eqnarray}
where the extra-dimensional parts are separately integrated:
\begin{eqnarray}
I_{1}^{n n^{\prime}}&=&\int d y d z W_{1}^{(n)} W_{1}^{(n ^{\prime})} ,\quad \quad \quad \quad
I_{2}^{n n^{\prime}}=\frac{1}{q+1} \int d y d z \partial_{y}\left(W_{1}^{(n)} e^{a A}\right) \partial_{y}\left(W_{1}^{(n ^{\prime})} e^{a A}\right) e^{-2 a A},\nonumber\\
I_{3}^{n n^{\prime}}&=&\frac{q^{2}}{q+1} \int d y d z W_{2}^{(n)} W_{2}^{(n ^{\prime})} ,\quad \quad
I_{4}^{n n^{\prime}}=\frac{1}{q+1} \int d y d z \partial_{z}\left(W_{1}^{(n)} e^{a A}\right) \partial_{z}\left(W_{1}^{(n ^{\prime})} e^{a A}\right) e^{-2 a A},\nonumber\\
I_{5}^{n n^{\prime}}&=&\frac{q^{2}}{q+1} \int d y d z W_{3}^{(n)} W_{3}^{\left(n^{\prime}\right)},\quad \quad
I_{6}^{n n^{\prime}}=\frac{(-1)^{q} q}{q+1} \int d y d z W_{2}^{(n)} \partial_{y}\left(W_{1}^{(n ^{\prime})} e^{a A}\right) e^{-a A},\nonumber\\
I_{7}^{n n^{\prime}} &=&\frac{q}{q+1} \int d y d z \partial_{z}\left(W_{2}^{(n)} e^{a A}\right) \partial_{z}\left(W_{2}^{\left(n^{\prime}\right)} e^{a A}\right) e^{-2 a A},\quad I_{8}^{n n^{\prime}}=\frac{(-1)^{q} q}{q+1} \int d y d z W_{3}^{(n)} \partial_{z}\left(W_{1}^{(n ^{\prime})} e^{a A}\right) e^{-a A},\nonumber\\
 I_{9}^{n n^{\prime}} &=&\frac{q}{q+1} \int d y d z \partial_{y}\left(W_{3}^{(n)} e^{a A}\right) \partial_{y}\left(W_{3}^{\left(n^{\prime}\right)} e^{a A}\right) e^{-2 a A},\nonumber\\
  I_{10}^{n n^{\prime}}&=&\frac{-q}{q+1} \int d y d z \partial_{y}\left(W_{3}^{(n)} e^{a A}\right) \partial_{z}\left(W_{2}^{\left(n^{\prime}\right)} e^{a A}\right) e^{-2 a A},\nonumber\\
I_{11}^{n n^{\prime}}&=&\frac{q(q-1)^{2}}{q+1} \int d y d z W_{4}^{(n)} W_{4}^{\left(n^{\prime}\right)},\quad  I_{12}^{n n^{\prime}}=\frac{(-1)^{q} q(q-1)}{q+1} \int d y d z \partial_{z}\left(W_{2}^{(n)} e^{a A}\right) W_{4}^{(n ^{\prime})} e^{-a A},\nonumber\\
I_{13}^{n n^{\prime}}&=&\frac{(-1)^{q-1} q(q-1)}{q+1} \int d y d z \partial_{y}\left(W_{3}^{(n)} e^{a A}\right) W_{4}^{(n ^{\prime})} e^{-a A}.
\label{I}
\end{eqnarray}
Here these parameters correspond to mass terms or coupling strength of fields on the brane.
 We impose the following orthogonality and finiteness conditions for a localizable case:
\begin{eqnarray}
&I_{1}^{nn^{\prime}}=\delta_{nn^{\prime}} ,\quad I_{3}^{nn^{\prime}}=I_{5}^{nn^{\prime}}=(q+1) \delta_{nn^{\prime}}, \quad I_{11}^{nn^{\prime}}=q(q+1) \delta_{nn^{\prime}},\nonumber \\&I_{k}^{n n^{\prime}}<\infty \quad {\rm otherwise.} \label{orthogonality condition}
\end{eqnarray}

In order to separate the EoMs into brane parts and extra-dimensional parts, we  will derive the EoMs in two different ways by exchanging the order of variation and KK decomposition. Since basically these two sets of EoMs should be consistent, we can get useful results by comparing them.
Firstly, we derive the EoMs by variating the effective action (\ref{effectiveaction}) with respect to the four types of fields $X_{(n)}$ , $\overline{X}_{(n)}$, $\underline{X}_{(n)}$ and $\overline{\underline{X}}_{(n)}$:
\begin{eqnarray}
 && \sum_{n^{\prime}}\left[\frac{1}{\sqrt{-\hat{g}}} \partial_{\mu}\left(\sqrt{-\hat{g}} I_{1}^{n n^{\prime}} Y_{\left(n^{\prime}\right)}^{\mu \mu_{1} \cdots \mu_{q}}\right)-\left(I_{2}^{n n^{\prime}}+I_{4}^{n n^{\prime}}\right) X_{\left(n^{\prime}\right)}^{\mu_{1} \cdots \mu_{q}}-I_{6}^{n n^{\prime}} \overline{Y}_{\left(n^{\prime}\right)}^{\mu_{1} \cdots \mu_{q}} -I_{8}^{n n^{\prime}} \underline{Y}_{\left(n^{\prime}\right)}^{\mu_{1} \cdots \mu_{q}}\right]=0,\nonumber \\
 && \sum_{n^{\prime}}\left[\frac{1}{\sqrt{-\hat{g}}} \partial_{\mu}\left(\sqrt{-\hat{g}} I_{3}^{n n^{\prime}} \overline{Y}_{\left(n^{\prime}\right)}^{\mu \mu_{1} \cdots \mu_{q-1}}+\sqrt{-\hat{g}} I_{6}^{n n^{\prime}} X_{\left(n^{\prime}\right)}^{\mu \mu_{1} \cdots \mu_{q-1}}\right)-I_{7}^{nn^{\prime}}\overline{X}_{\left(n^{\prime}\right)}^{\mu_{1} \cdots \mu_{q-1}} -I_{10}^{n n^{\prime}} \underline{X}_{\left(n^{\prime}\right)}^{\mu_{1} \cdots \mu_{q-1}}-I_{12}^{n n^{\prime}} \overline{\underline{Y}}_{\left(n^{\prime}\right)}^{\mu_{1} \cdots \mu_{q-1}}\right]=0 ,\nonumber\\
  && \sum_{n^{\prime}}\left[\frac{1}{\sqrt{-\hat{g}}} \partial_{\mu}\left(\sqrt{-\hat{g}} I_{5}^{n n^{\prime}} \underline{Y}_{\left(n^{\prime}\right)}^{\mu \mu_{1} \cdots \mu_{q-1}}+\sqrt{-\hat{g}} I_{8}^{n n^{\prime}} X_{\left(n^{\prime}\right)}^{\mu \mu_{1} \cdots \mu_{q-1}}\right)-I_{10}^{nn^{\prime}}\overline{X}_{\left(n^{\prime}\right)}^{\mu_{1} \cdots \mu_{q-1}} -I_{9}^{n n^{\prime}} \underline{X}_{\left(n^{\prime}\right)}^{\mu_{1} \cdots \mu_{q-1}}-I_{13}^{n n^{\prime}} \overline{\underline{Y}}_{\left(n^{\prime}\right)}^{\mu_{1} \cdots \mu_{q-1}}\right]=0, \nonumber\\
    && \sum_{n^{\prime}}\frac{1}{\sqrt{-\hat{g}}} \partial_{\mu}\left[{\sqrt{-\hat{g}}}\left(I_{11}^{n n^{\prime} }\overline{\underline{Y}}_{\left(n^{\prime}\right)}^{\mu\mu_{1} \cdots \mu_{q-2}} +I_{12}^{n n^{\prime} }\overline{X}_{\left(n^{\prime}\right)}^{\mu\mu_{1} \cdots \mu_{q-2}}
    +I_{13}^{n n^{\prime} }\underline{X}_{\left(n^{\prime}\right)}^{\mu\mu_{1} \cdots \mu_{q-2}}
    \right)\right]=0. \nonumber \\ \label{eom1}
\end{eqnarray}
Alternatively, we can obtain EoMs by directly inserting the decomposition (4) into the undecomposed EoMs (2)
\begin{eqnarray}
&&\frac {1}{\sqrt{-\hat{g}}}\partial_{\mu_{1}}\left(\sqrt{-\hat{g}}Y_{(n)}^{\mu_{1}\cdots\mu_{q+1}}\right)+\left(\lambda_{1}^{(n)}+\lambda_{2}^{(n)}\right)X_{(n)}^{\mu_{2}\cdots\mu_{q+1}}+\lambda_{3}^{(n)}\overline{Y}_{(n)}^{\mu_{2}\cdots\mu_{q+1}}+\lambda_{4}^{(n)}\underline{Y}_{(n)}^{\mu_{2}\cdots\mu_{q+1}}=0,\nonumber\\
&&\frac {1}{\sqrt{-\hat{g}}}\partial_{\mu_{1}}\left(\sqrt{-\hat{g}}\overline{Y}_{(n)}^{\mu_{1}\cdots\mu_{q}}\right)+\frac {\lambda_{5}^{(n)}}{\sqrt{-\hat{g}}}\partial_{\mu_{1}}\left(\sqrt{-\hat{g}}X_{(n)}^{\mu_{1}\cdots\mu_{q}}\right)+\lambda_{6}^{(n)}\underline{X}_{(n)}^{\mu_{2}\cdots\mu_{q}}+\lambda_{7}^{(n)}\overline{X}_{(n)}^{\mu_{2}\cdots\mu_{q}}+\lambda_{8}^{(n)}\overline{\underline{Y}}_{(n)}^{\mu_{2}\cdots\mu_{q}}=0,\nonumber\\
&&\frac {1}{\sqrt{-\hat{g}}}\partial_{\mu_{1}}\left(\sqrt{-\hat{g}}\underline{Y}_{(n)}^{\mu_{1}\cdots\mu_{q}}\right)+\frac {\lambda_{9}^{(n)}}{\sqrt{-\hat{g}}}\partial_{\mu_{1}}\left(\sqrt{-\hat{g}}X_{(n)}^{\mu_{1}\cdots\mu_{q}}\right)+\lambda_{10}^{(n)}\underline{X}_{(n)}^{\mu_{2}\cdots\mu_{q}}+\lambda_{11}^{(n)}\overline{X}_{(n)}^{\mu_{2}\cdots\mu_{q}}+\lambda_{12}^{(n)}\overline{\underline{Y}}_{(n)}^{\mu_{2}\cdots\mu_{q}}=0,\nonumber\\
&&\frac {1}{\sqrt{-\hat{g}}}\partial_{\mu_{1}}\left[\sqrt{-\hat{g}}\left(\overline{\underline{Y}}_{(n)}^{\mu_{1}\cdots\mu_{q-1}}+\lambda_{13}^{(n)}\overline{X}_{(n)}^{\mu_{1}\cdots\mu_{q-1}}+\lambda_{14}^{(n)}\underline{X}_{(n)}^{\mu_{1}\cdots\mu_{q-1}}\right)\right]=0, \nonumber\\ \label{eom2}
\end{eqnarray}
where
\begin{eqnarray}
&&\lambda_{1}^{(n)}=\frac{\partial_{y}\left[\partial_{y}\left(W_{1}^{(n)}e^{aA}\right)e^{-2aA}\right]}{e^{-aA}\left(q+1\right)W_{1}^{(n)}},\;
\lambda_{2}^{(n)}=\frac{\partial_{z}\left[\partial_{z}\left(W_{1}^{(n)}e^{aA}\right)e^{-2aA}\right]}{e^{-aA}\left(q+1\right)W_{1}^{(n)}},\;
\lambda_{3}^{(n)}=\frac{(-1)^{q}q\partial_{y}\left(W_{2}^{(n)}e^{-aA}\right)}{e^{-aA}\left(q+1\right)W_{1}^{(n)}},\nonumber\\
&&\lambda_{4}^{(n)}=\frac{(-1)^{q}q\partial_{z}\left(W_{3}^{(n)}e^{-aA}\right)}{e^{-aA}\left(q+1\right)W_{1}^{(n)}},\quad
\lambda_{5}^{(n)}=\frac{(-1)^{q}\partial_{y}\left(W_{1}^{(n)}e^{aA}\right)}{e^{aA}qW_{2}^{(n)}},\quad
\quad\lambda_{6}^{(n)}=\frac{\partial_{z}\left[\partial_{y}\left(W_{3}^{(n)}e^{aA}\right)e^{-2aA}\right]}{-qe^{-aA}W_{2}^{(n)}},\nonumber\\
&&\lambda_{7}^{(n)}=\frac{\partial_{z}\left[\partial_{z}\left(W_{2}^{(n)}e^{aA}\right)e^{-2aA}\right]}{e^{-aA}qW_{2}^{(n)}},\;
\lambda_{8}^{(n)}=\frac{(q-1)\partial_{z}\left(W_{4}^{(n)}e^{-aA}\right)}{(-1)^{q}e^{-aA}qW_{2}^{(n)}},\quad
\lambda_{9}^{(n)}=\frac{\partial_{z}\left(W_{1}^{(n)}e^{aA}\right)}{(-1)^{q}qe^{aA}W_{3}^{(n)}},\nonumber\\
&&\lambda_{10}^{(n)}=\frac{\partial_{y}\left[\partial_{y}\left(W_{3}^{(n)}e^{aA}\right)e^{-2aA}\right]}{e^{-aA}qW_{3}^{(n)}},\;
\lambda_{11}^{(n)}=\frac{\partial_{y}\left[\partial_{z}\left(W_{2}^{(n)}e^{aA}\right)e^{-2aA}\right]}{-qe^{-aA}W_{3}^{(n)}},
\;\lambda_{12}^{(n)}=\frac{(q-1)\partial_{y}\left(W_{4}^{(n)}e^{-aA}\right)}{(-1)^{q-1}qe^{-aA}W_{3}^{(n)}},\nonumber\\
&&\lambda_{13}^{(n)}=\frac{(-1)^{q}\partial_{z}\left(W_{2}^{(n)}e^{aA}\right)}{\left(q-1\right)e^{aA}W_{4}^{(n)}},\quad\quad
\lambda_{14}^{(n)}=\frac{(-1)^{q-1}\partial_{y}\left(W_{3}^{(n)}e^{aA}\right)}{\left(q-1\right)e^{aA}W_{4}^{(n)}}\label{lmd}.
\end{eqnarray}
By comparing Eq.(\ref{eom1}) with  Eq.(\ref{eom2}), we find the $\lambda_i^{(n)}$s are necessarily constants under our former assumption, so we introduce the following mass parameters
\begin{equation}
\lambda_{1}^{(n)}=\frac{-m_{1y}^{(n)2}}{q+1},\quad \quad
\lambda_{2}^{(n)}=\frac{-m_{1z}^{(n)2}}{q+1},\quad \quad
\lambda_{7}^{(n)}=\frac{-m_{2z}^{(n)2}}{q},\quad \quad
\lambda_{10}^{(n)}=\frac{-m_{3y}^{(n)2}}{q},
\end{equation}
which give a set of Schr\"{o}dinger-like equations for the extra-dimensional functions
\begin{eqnarray}
&&\left[-\partial_{z}^{2}+(a\partial_{z}A)^{2}-a\partial_{z}^{2}A\right]W_{1}^{(n)}=m_{1z}^{(n)2}W_{1}^{(n)},\quad
\left[-\partial_{y}^{2}+(a\partial_{y}A)^{2}-a\partial_{y}^{2}A\right]W_{1}^{(n)}=m_{1y}^{(n)2}W_{1}^{(n)},\nonumber\\
&&\left[-\partial_{z}^{2}+(a\partial_{z}A)^{2}-a\partial_{z}^{2}A\right]W_{2}^{(n)}=m_{2z}^{(n)2}W_{2}^{(n)},\quad
\left[-\partial_{y}^{2}+(a\partial_{y}A)^{2}-a\partial_{y}^{2}A\right]W_{3}^{(n)}=m_{3y}^{(n)2}W_{3}^{(n)}.
\end{eqnarray}
Notice Eq.(\ref{orthogonality condition}) and compare the EoMs again to get the relations
\begin{equation}
\lambda_{3}^{(n)}=-(q+1)\lambda_{5}^{(n)},\quad
\lambda_{4}^{(n)}=-(q+1)\lambda_{9}^{(n)},\quad
\lambda_{6}^{(n)}=\lambda_{11}^{(n)},\quad
\lambda_{12}^{(n)}=-q\lambda_{14}^{(n)},\quad
\lambda_{8}^{(n)}=-q\lambda_{13}^{(n)}.\\
\end{equation}
Substitution of $\lambda_{3}^{(n)}$ into $\lambda_{5}^{(n)}$, $\lambda_{4}^{(n)}$ into $\lambda_{9}^{(n)}$, $\lambda_{14}^{(n)}$ into $\lambda_{12}^{(n)}$, $\lambda_{13}^{(n)}$ into $\lambda_{8}^{(n)}$ yields
\begin{eqnarray}
&&\lambda_{3}^{(n)}\lambda_{5}^{(n)}=\lambda_{1}^{(n)},\quad\quad
\lambda_{3}^{(n)}=m_{1y}^{(n)},\quad\quad\lambda_{5}^{(n)}=\frac{-m_{1y}^{(n)}}{q+1},\nonumber\\
&&\lambda_{4}^{(n)}\lambda_{9}^{(n)}=\lambda_{2}^{(n)},\quad\quad
\lambda_{4}^{(n)}=m_{1z}^{(n)},\quad\quad\lambda_{9}^{(n)}=\frac{-m_{1z}^{(n)}}{q+1},\nonumber\\
&&\lambda_{12}^{(n)}\lambda_{14}^{(n)}=\lambda_{10}^{(n)},\quad\quad
\lambda_{12}^{(n)}=m_{3y}^{(n)},\quad\quad\lambda_{14}^{(n)}=\frac{-m_{3y}^{(n)}}{q},\nonumber\\
&&\lambda_{8}^{(n)}\lambda_{13}^{(n)}=\lambda_{7}^{(n)},\quad\quad
\lambda_{8}^{(n)}=-m_{2z}^{(n)},\quad\;\lambda_{13}^{(n)}=\frac{m_{2z}^{(n)}}{q},\nonumber\\
\end{eqnarray}
where we have chosen signs for later consistence.

Similarly, by substituting $\lambda_{5}^{(n)}$ into $\lambda_{3}^{(n)}$, $\lambda_{9}^{(n)}$ into $\lambda_{4}^{(n)}$, $\lambda_{12}^{(n)}$ into $\lambda_{14}^{(n)}$, $\lambda_{8}^{(n)}$ into $\lambda_{13}^{(n)}$, we obtain another four  Schr\"{o}dinger-like  equations with mass terms
\begin{eqnarray}
&&\left[-\partial_{y}^{2}+(a\partial_{y}A)^{2}+a\partial_{y}^{2}A\right]W_{2}^{(n)}=m_{1y}^{(n)2}W_{2}^{(n)},\quad
\left[-\partial_{z}^{2}+(a\partial_{z}A)^{2}+a\partial_{z}^{2}A\right]W_{3}^{(n)}=m_{1z}^{(n)2}W_{3}^{(n)},\quad \nonumber\\
&&\left[-\partial_{y}^{2}+(a\partial_{y}A)^{2}+a\partial_{y}^{2}A\right]W_{4}^{(n)}=m_{3y}^{(n)2}W_{4}^{(n)},\quad
\left[-\partial_{z}^{2}+(a\partial_{z}A)^{2}+a\partial_{z}^{2}A\right]W_{4}^{(n)}=m_{2z}^{(n)2}W_{4}^{(n)}.
\end{eqnarray}
Comparing the expressions of $\lambda_{5}$, $\lambda_{9}$, $\lambda_{6}$, and those of $\lambda_{5}$, $\lambda_{9}$, $\lambda_{11}$, we have
\begin{equation}
\lambda_{5}^{(n)}\lambda_{7}^{(n)}=-\lambda_{6}^{(n)}\lambda_{9}^{(n)},\quad\quad
\lambda_{9}^{(n)}\lambda_{10}^{(n)}=-\lambda_{5}^{(n)}\lambda_{11}^{(n)},\quad\quad
\lambda_{6}^{(n)}=\frac{m_{1y}^{(n)}m_{2z}^{(n)2}}{qm_{1z}^{(n)}},
\end{equation}
which lead to $\frac{m_{3y}^{(n)2}}{m_{2z}^{(n)2}}=\frac{m_{1y}^{(n)2}}{m_{1z}^{(n)2}}$, therefore we can introduce constant parameters $\eta^{(n)}$s such that $m_{2z}^{(n)}=\eta^{(n)}m_{1z}^{(n)}$, $m_{3y}^{(n)}=\eta^{(n)}m_{1y}^{(n)}$, and rewrite
$m_{1y}^{(n)}$, $m_{1z}^{(n)}$ as $m_{y}^{(n)}$, $m_{z}^{(n)}$.
Consequently we can summarize the eight Schr\"{o}dinger-like equations as
\begin{eqnarray}
&&\left[-\partial_{y}^{2}+(a\partial_{y}A)^{2}-a\partial_{y}^{2}A\right]W_{1}^{(n)}=m_{y}^{(n)2}W_{1}^{(n)},\quad\quad\;\;
\left[-\partial_{z}^{2}+(a\partial_{z}A)^{2}-a\partial_{z}^{2}A\right]W_{1}^{(n)}=m_{z}^{(n)2}W_{1}^{(n)},\nonumber\\
&&\left[-\partial_{y}^{2}+(a\partial_{y}A)^{2}+a\partial_{y}^{2}A\right]W_{2}^{(n)}=m_{y}^{(n)2}W_{2}^{(n)},\quad\quad\;\;
\left[-\partial_{z}^{2}+(a\partial_{z}A)^{2}-a\partial_{z}^{2}A\right]W_{2}^{(n)}=\eta_{(n)}^{2}m_{z}^{(n)2}W_{2}^{(n)},\nonumber\\
&&\left[-\partial_{y}^{2}+(a\partial_{y}A)^{2}-a\partial_{y}^{2}A\right]W_{3}^{(n)}=\eta_{(n)}^{2}m_{y}^{(n)2}W_{3}^{(n)},\quad\;
\left[-\partial_{z}^{2}+(a\partial_{z}A)^{2}+a\partial_{z}^{2}A\right]W_{3}^{(n)}=m_{z}^{(n)2}W_{3}^{(n)},\nonumber\\
&&\left[-\partial_{y}^{2}+(a\partial_{y}A)^{2}+a\partial_{y}^{2}A\right]W_{4}^{(n)}=\eta_{(n)}^{2}m_{y}^{(n)2}W_{4}^{(n)},\quad\;
\left[-\partial_{z}^{2}+(a\partial_{z}A)^{2}+a\partial_{z}^{2}A\right]W_{4}^{(n)}=\eta_{(n)}^{2}m_{z}^{(n)2}W_{4}^{(n)}.\nonumber\\
\end{eqnarray}
From the above equations we obtain two series of mass parameters form the two codimensions, and mass spectra of each codimension are quite similar, in spite of common factors $\eta_{(n)}$s. Under certain background solutions such that $\partial_y\partial_zA=0$, or the condition that one of $W_2^{(n)}$ and $W_3^{(n)}$ is zero, we will have $\eta_{(n)}=1$. As usual, the operator on the left side of each equation can be written as a product of an operator and its conjugate, $PP^\dagger$, therefore mass spectra of the KK modes are nonnegative definite. With an explicit solution of the space-time background, one can solve these equations to get their mass spectra and wave functions. Since the subsequent discussion is solution-independent, we do not analyse these equations further and reserve localizability as an assumption.

\section{Gauge invariance of the brane action}\label{invariance}

With the previous results, one can continue comparing the two sets of EoMs to exhaust relations between the $I^{nn{\prime}}$s and the $m^{(n)}$s
\begin{eqnarray}
&&I_{1}^{nn^{\prime}}=\delta_{nn^{\prime}},\quad\quad
I_{3}^{nn^{\prime}}=I_{5}^{nn^{\prime}}=(q+1)\delta_{nn^{\prime}},\quad\quad
I_{11}^{nn^{\prime}}=q(q+1)\delta_{nn^{\prime}},\nonumber\\
&&I_{2}^{nn^{\prime}}+I_{4}^{nn^{\prime}}=-\left(\lambda_{1}^{(n)}+\lambda_{2}^{(n)}\right)\delta_{nn^{\prime}}=\frac{\delta_{nn^{\prime}}}{q+1}\left(m_{y}^{(n)2}+m_{z}^{(n)2}\right),\quad\quad
I_{6}^{nn^{\prime}}=-\lambda_{3}^{(n)}\delta_{nn^{\prime}}=-m_{y}^{(n)}\delta_{nn^{\prime}},\nonumber\\
&&I_{7}^{nn^{\prime}}=-(q+1)\lambda_{7}^{(n)}\delta_{nn^{\prime}}=\frac{q+1}{q}\eta_{(n)}^{2}m_{z}^{(n)2}\delta_{nn^{\prime}},\quad\quad
I_{8}^{nn^{\prime}}=-\lambda_{4}^{(n)}\delta_{nn^{\prime}}=-m_{z}^{(n)}\delta_{nn^{\prime}},\nonumber\\
&&I_{9}^{nn^{\prime}}=-(q+1)\lambda_{10}^{(n)}\delta_{nn^{\prime}}=\frac{q+1}{q}\eta_{(n)}^{2}m_{y}^{(n)2}\delta_{nn^{\prime}},\quad\quad
I_{10}^{nn^{\prime}}=-(q+1)\lambda_{6}^{(n)}\delta_{nn^{\prime}}=\frac{q+1}{-q}\eta_{(n)}^{2}m_{y}^{(n)}m_{z}^{(n)}\delta_{nn^{\prime}},\nonumber\\
&&I_{12}^{nn^{\prime}}=-(q+1)\lambda_{8}^{(n)}\delta_{nn^{\prime}}=(q+1)\eta_{(n)}m_{z}^{(n)}\delta_{nn^{\prime}},\quad\quad
I_{13}^{nn^{\prime}}=-(q+1)\lambda_{12}^{(n)}\delta_{nn^{\prime}}=-(q+1)\eta_{(n)}m_{y}^{(n)}\delta_{nn^{\prime}}.\nonumber\\ \label{I2m}
\end{eqnarray}
By pluging these relations into Eq.(\ref{effectiveaction}), we can convert the effective action on the brane into
\begin{eqnarray}
S=&&\frac{-1}{2(q+1)!}\sum_{n}\int d^{p+1}x\sqrt{-\hat{g}}
\left[\left(Y^{(n)}_{\mu_{1}\cdots\mu_{q+1}}\right)^{2}+(q+1)\left(\overline{Y}^{(n)}_{\mu_{1}\cdots\mu_{q}}\right)^{2}+
(q+1)\left(\underline{Y}^{(n)}_{\mu_{1}\cdots\mu_{q}}\right)^{2}\right.\nonumber\\
&&+q(q+1)\left(\underline{\overline{Y}}^{(n)}_{\mu_{1}\cdots\mu_{q-1}}\right)^{2}
+q(q+1)\left(\underline{\overline{Y}}^{(n)}_{\mu_{1}\cdots\mu_{q-1}}\right)^{2}      +\frac{1}{q+1}\left(m_{y}^{(n)2}+m_{z}^{(n)2}\right)\left(X^{(n)}_{\mu_{1}\cdots\mu_{q}}\right)^{2}\nonumber\\
&&-2m_{y}^{(n)}X_{(n)}^{\mu_{1}\cdots\mu_{q}}\overline{Y}^{(n)}_{\mu_{1}\cdots\mu_{q}}
-2m_{z}^{(n)}X_{(n)}^{\mu_{1}\cdots\mu_{q}}\underline{Y}^{(n)}_{\mu_{1}\cdots\mu_{q}}
+\frac{q+1}{q}\eta_{(n)}^{2}m_{z}^{(n)2}\left(\overline{X}^{(n)}_{\mu_{1}\cdots\mu_{q-1}}\right)^{2}\nonumber\\
&&+\frac{q+1}{q}\eta_{(n)}^{2}m_{y}^{(n)2}\left(\underline{X}^{(n)}_{\mu_{1}\cdots\mu_{q-1}}\right)^{2}
-\frac{2(q+1)}{q}\eta_{(n)}^{2}m_{y}^{(n)}m_{z}^{(n)}\overline{X}^{(n)}_{\mu_{1}\cdots\mu_{q-1}}\underline{X}_{(n)}^{\mu_{1}\cdots\mu_{q-1}}\nonumber\\
&&\left.+2(q+1)\eta_{(n)}m_{z}^{(n)}\overline{X}^{(n)}_{\mu_{1}\cdots\mu_{q-1}}\overline{\underline{Y}}_{(n)}^{\mu_{1}\cdots\mu_{q-1}}
-2(q+1)\eta_{(n)}m_{Y}^{(n)}\overline{X}^{(n)}_{\mu_{1}\cdots\mu_{q-1}}\underline{\overline{Y}}_{(n)}^{\mu_{1}\cdots\mu_{q-1}} \right]\nonumber\\
=&&\frac{-1}{2(q+1)!}\sum_{n}\int d^{p+1}x\sqrt{-\hat{g}}
\left(Y^{(n)}_{\mu_{1}\cdots\mu_{q+1}}\right)^{2}+\nonumber\\
&&\frac{-1}{2q!}\sum_{n}\int d^{p+1}x\sqrt{-\hat{g}}
\left[\left(\overline{Y}^{(n)}_{\mu_{1}\cdots\mu_{q}}-\frac{m_{y}^{(n)}}{q+1}X^{(n)}_{\mu_{1}\cdots\mu_{q}}\right)^{2}
+\left(\underline{Y}^{(n)}_{\mu_{1}\cdots\mu_{q}}-\frac{m_{z}^{(n)}}{q+1}X^{(n)}_{\mu_{1}\cdots\mu_{q}}\right)^{2}\right]+\nonumber\\
&&\frac{-1}{2(q-1)!}\sum_{n}\int d^{p+1}x\sqrt{-\hat{g}}
\left(\underline{\overline{Y}}^{(n)}_{\mu_{1}\cdots\mu_{q-1}}
+\frac{\eta_{(n)}m_{z}^{(n)}}{q}\overline{X}^{(n)}_{\mu_{1}\cdots\mu_{q-1}}
-\frac{\eta_{(n)}m_{y}^{(n)}}{q}\underline{X}^{(n)}_{\mu_{1}\cdots\mu_{q-1}}
\right)^{2}. \label{braneaction}
\end{eqnarray}

One can observe that through coupling of the four fields, the effective action appears exactly as a sum of four squares, which is then apparently invariant under the following gauge transformation
\begin{eqnarray}
&X^{(n)}_{\mu_{1}\cdots\mu_{q}}\rightarrow X^{(n)}_{\mu_{1}\cdots\mu_{q}}+\partial_{[\mu_{1}}\Lambda^{(n)}_{\mu_{2}\cdots\mu_{q}]} , \nonumber\\
&\overline{X}^{(n)}_{\mu_{1}\cdots\mu_{q-1}}\rightarrow \overline{X}^{(n)}_{\mu_{1}\cdots\mu_{q-1}}+\frac{m^{(n)}_{y}}{q+1}\Lambda^{(n)}_{\mu_{1}\cdots\mu_{q-1}},\nonumber\\
&\underline{X}^{(n)}_{\mu_{1}\cdots\mu_{q-1}}\rightarrow
\underline{X}^{(n)}_{\mu_{1}\cdots\mu_{q-1}}+\frac{m_{z}^{(n)}}{q+1}\Lambda^{(n)}_{\mu_{1}\cdots\mu_{q-1}} ,\nonumber\\
&\underline{\overline{X}}^{(n)}_{\mu_{1}\cdots\mu_{q-2}}\rightarrow
\underline{\overline{X}}^{(n)}_{\mu_{1}\cdots\mu_{q-2}}+\partial_{[\mu_{1}}\Gamma^{(n)}_{\mu_{2}\cdots\mu_{q-2}]},
\end{eqnarray}
where $\Lambda^{(n)}$ and $\Gamma^{(n)}$ are arbitrary $(q-1)$-form and $(q-2)$-form, respectively.

\section{Induced Hodge dualities on the brane}\label{Hodge dualities}
In the bulk, a $q$-form potential $X$ is said to be dual  to a $(D-2-q)$-form potential $X^{*}$ (though not uniquely determined) via Hodge duality between their strength fields $Y$ and $Y^{*}$
\begin{equation}
Y^{*M_{1}\cdots M_{p+2-q}}=\frac{\epsilon^{M_{1}\cdots M_{p+2-q}N_{1}\cdots N_{q+1}}}{(q-1)!\sqrt{-g}}Y_{N_{1}\cdots N_{q+1}}, \label{strength fields}
\end{equation}
i.e, a $q+1 \sim p+2-q$ duality. A direct substitution  into  Eq. (\ref{action}) shows
\begin{eqnarray}
S&=&-\frac{1}{2(q+1)!}\int d^{D}x\sqrt{-g}Y^{N_{1}\cdots N_{q+1}}Y_{N_{1}\cdots N_{q+1}} \nonumber\\
&=&-\frac{1}{2(p+2-q)!}\int d^{D}x\sqrt{-g}Y^{*N_{1}\cdots N_{p+2-q}}Y_{N_{1}\cdots N_{p+2-q}}^{*}=S^{*},
\end{eqnarray}
which indicates two massless dual potentials $X$ and $X^{*}$ are physically equivalent.

However, when performing localization for $X$ and $X^{*}$ through the usual KK decomposition, we get two brane fields of the same order as their bulk ones, which are therefore impossible to be Hodge-dual on the brane, for dimensional reason. On the other hand, the new decomposition mechanism yields a series of brane forms of different orders, and the dimensional requirement is satisfied when we match them reversely. Therefore it is very likely that the bulk duality will naturally reduce to brane dualities, as what we are to show.

In what follows we denote $X^{*}_{(n)}$ and $Y^{*}_{(n)}$ as the brane components of $X^{*}$ and  $Y^{*}$ rather than the duals of some fields. From Eq. (\ref{strength fields}), one can obtain
\begin{eqnarray}
\sqrt{-g}Y^{*\mu_{1}\cdots\mu_{p+2-q}}=\frac{\epsilon^{\mu_{1}\cdots\mu_{p+2-q}\nu_{1}\cdots\nu_{q-1}yz}}{(q-1)!}Y_{\nu_{1}\cdots\nu_{q-1}yz},\quad
\sqrt{-g}Y^{*\mu_{1}\cdots\mu_{p+1-q}y}=\frac{\epsilon^{\mu_{1}\cdots\mu_{p+1-q}y\nu_{1}\cdots\nu_{q}z}}{q!}Y_{\nu_{1}\cdots\nu_{q}z},\nonumber\\
\sqrt{-g}Y^{*\mu_{1}\cdots\mu_{p+1-q}z}=\frac{\epsilon^{\mu_{1}\cdots\mu_{p+1-q}z\nu_{1}\cdots\nu_{q}y}}{q!}Y_{\nu_{1}\cdots\nu_{q}y} ,\quad
\sqrt{-g}Y^{*\mu_{1}\cdots\mu_{p-q}yz}=\frac{\epsilon^{\mu_{1}\cdots\mu_{p-q}yz\nu_{1}\cdots\nu_{q+1}}}{(q+1)!}Y_{\nu_{1}\cdots\nu_{q+1}} \label{strength fields2}.
\end{eqnarray}
Notice that when working with the dual fields we just need to replace $q+1$ by $p+2-q$. So the substitution of (\ref{field strength1}) into the first equation of Eq. (\ref{strength fields2})  gives
\begin{eqnarray}
&\sqrt{-g}&\sum_{n}Y_{(n)}^{*\mu_{1}\cdots\mu_{p+2-q}}W_{1}^{*(n)}e^{-(a+(p+2-q))A}\nonumber\\
&&=\sum_{n}\frac{\epsilon^{\mu_{1}\cdots\mu_{p+2-q}\nu_{1}\cdots\nu_{q-1}yz}}{(q-1)!}
\left[\frac{(-1)^{q-1}}{q+1}\underline{X}^{(n)}_{\mu_{1}\cdots\mu_{q-1}}\partial_{y}\left(W_{3}^{(n)}e^{aA}\right)
+\frac{(-1)^{q}}{q+1}\overline{X}^{(n)}_{\mu_{1}\cdots\mu_{q-1}}\partial_{z}\left(W_{2}^{(n)}e^{aA}\right)\right.\nonumber\\
&&\quad\left.+\frac{q-1}{q+1}\underline{\overline{Y}}^{(n)}_{\mu_{1}\cdots\mu_{q-1}}W_{4}^{(n)}e^{aA}
\right]\nonumber\\
&&=\sum_{n}\frac{\epsilon^{\mu_{1}\cdots\mu_{p+2-q}\nu_{1}\cdots\nu_{q-1}yz}}{(q-1)!}
\left[\frac{q-1}{q+1}\lambda_{14}^{(n)}W_{4}^{(n)}e^{aA}\underline{X}^{(n)}_{\mu_{1}\cdots\mu_{q-1}}
+\frac{q-1}{q+1}\lambda_{13}^{(n)}W_{4}^{(n)}e^{aA}\overline{X}^{(n)}_{\mu_{1}\cdots\mu_{q-1}}\right.\nonumber\\
&&\quad\left.+\frac{q-1}{q+1}\underline{\overline{Y}}^{(n)}_{\mu_{1}\cdots\mu_{q-1}}W_{4}^{(n)}e^{aA}
\right]\nonumber\\
&&=\frac{\epsilon^{\mu_{1}\cdots\mu_{p+2-q}\nu_{1}\cdots\nu_{q-1}yz}W_{4}^{(n)}e^{aA}}{(q-2)!(q+1)}
\left(\underline{\overline{Y}}^{(n)}_{\mu_{1}\cdots\mu_{q-1}}
+\frac{\eta_{(n)}m_{z}^{(n)}}{q}\overline{X}^{(n)}_{\mu_{1}\cdots\mu_{q-1}}
-\frac{\eta_{(n)}m_{y}^{(n)}}{q}\underline{X}^{(n)}_{\mu_{1}\cdots\mu_{q-1}}
\right),
\end{eqnarray}
which can be simplified to
\begin{equation}
\sqrt{-\hat{g}}\sum_{n}Y_{(n)}^{*\mu_{1}\cdots\mu_{p+2-q}}W_{1}^{*(n)}
=\sum_{n}\frac{\epsilon^{\mu_{1}\cdots\mu_{p+2-q}\nu_{1}\cdots\nu_{q-1}}W_{4}^{(n)}}{(q-2)!(q+1)}
\left(\underline{\overline{Y}}^{(n)}_{\mu_{1}\cdots\mu_{q-1}}
+\frac{\eta_{(n)}m_{z}^{(n)}}{q}\overline{X}^{(n)}_{\mu_{1}\cdots\mu_{q-1}}
-\frac{\eta_{(n)}m_{y}^{(n)}}{q}\underline{X}^{(n)}_{\mu_{1}\cdots\mu_{q-1}}
\right).\label{Simplify}
\end{equation}
Here we find that $W_{1}^{*(n)}$ is proportional to $W_{4}^{(n)}$, so according to the relation
$\frac{q-1}{q+1}\int dydzW_{4}^{(n)}W_{4}^{(n^{\prime})}\\
=\delta_{nn^{\prime}}=\int dydzW_{1}^{*(n)}W_{1}^{*(n^{\prime})}$ , we can let $W_{4}^{(n)}=\frac{q+1}{q-1}W_{1}^{*(n)}$, then Eq. (\ref{Simplify}) is further simplified as
\begin{equation}
\sqrt{-\hat{g}}Y_{(n)}^{*\mu_{1}\cdots\mu_{p+2-q}}
=\frac{\epsilon^{\mu_{1}\cdots\mu_{p+2-q}\nu_{1}\cdots\nu_{q-1}}}{(q-1)!}
\left(\underline{\overline{Y}}^{(n)}_{\mu_{1}\cdots\mu_{q-1}}
+\frac{\eta_{(n)}m_{z}^{(n)}}{q}\overline{X}^{(n)}_{\mu_{1}\cdots\mu_{q-1}}
-\frac{\eta_{(n)}m_{y}^{(n)}}{q}\underline{X}^{(n)}_{\mu_{1}\cdots\mu_{q-1}}
\right).  \label{Hodge duality1}
\end{equation}
This is exactly a pair of Hodge duality on the brane.

Similarly, the remaining three equations of Eq. (\ref{strength fields2}) will lead to
\begin{eqnarray}
&&W_{3}^{(n)}=\frac{q+1}{q}\frac{p+1-q}{p+2-q}W_{2}^{*(n)},\nonumber\\
&&\sqrt{-\hat{g}}
\left(\overline{Y}_{(n)}^{*\mu_{1}\cdots\mu_{p+1-q}}-\frac{m_{y}^{*(n)}}{p+2-q}X_{(n)}^{*\mu_{1}\cdots\mu_{p+1-q}}\right)
=\frac{\epsilon^{\mu_{1}\cdots\mu_{p+1-q}\nu_{1}\cdots\nu_{q}}}{(-1)^{q}q!}
\left(\underline{Y}^{(n)}_{*\nu_{1}\cdots\nu_{q}}-\frac{m_{z}^{(n)}}{q+1}X^{(n)}_{*\nu_{1}\cdots\nu_{q}}\right);\nonumber\\
&&W_{2}^{(n)}=\frac{q+1}{q}\frac{p+1-q}{p+2-q}W_{3}^{*(n)},\nonumber\\
&&\sqrt{-\hat{g}}
\left(\underline{Y}_{(n)}^{*\mu_{1}\cdots\mu_{p+1-q}}-\frac{m_{z}^{*(n)}}{p+2-q}X_{(n)}^{*\mu_{1}\cdots\mu_{p+1-q}}\right)
=\frac{\epsilon^{\mu_{1}\cdots\mu_{p+1-q}\nu_{1}\cdots\nu_{q}}}{(-1)^{q+1}q!}
\left(\overline{Y}^{(n)}_{*\nu_{1}\cdots\nu_{q}}-\frac{m_{y}^{(n)}}{q+1}X^{(n)}_{*\nu_{1}\cdots\nu_{q}}\right);\nonumber\\
&&W_{1}^{(n)}=\frac{p-q}{p+2-q}W_{4}^{*(n)},\nonumber\\
&&\sqrt{-\hat{g}}
\left(\underline{\overline{Y}}_{(n)}^{*\mu_{1}\cdots\mu_{p-q}}
+\frac{\eta^*_{(n)}m_{z}^{*(n)}}{p+1-q}\overline{X}_{(n)}^{*\mu_{1}\cdots\mu_{p-q}}
-\frac{\eta^*_{(n)}m_{y}^{*(n)}}{p+1-q}\underline{X}_{(n)}^{*\mu_{1}\cdots\mu_{p-q}}\right)
=\frac{\epsilon^{\mu_{1}\cdots\mu_{p-q}\nu_{1}\cdots\nu_{q+1}}}{(q+1)!}Y^{(n)}_{\nu_{1}\cdots\nu_{q+1}} \label{Hodge duality2}.
\end{eqnarray}
These are another three pairs of dualities and relations between extra-dimensional functions,
 where, because of the correspondence between the extra-dimensional functions, the mass spectra for the dual modes are altered via $m_y^{*(n)}=\eta_{(n)}m_y^{(n)}$, $m_z^{*(n)}=\eta_{(n)}m_z^{(n)}$,  $\eta_{(n)}^*=1/\eta_{(n)}$. Incidentally, our $\pm$ sign choices are determined by the identity $**=(-1)^{sgn(g)+(n-k)k}$.

To sum up, the bulk duality of a ($q+1$)-form strength and a ($p+2-q$)-form strength (in other words, of potential ranks $q$ and $p+1-q$), generates four coupled dualities on the brane,
which is illustrated as follows: \\

~~~~~~~~~~~~~~~~~~~~~~~~~~~~\begin{tabular}{|c|c|}
	\hline
	& Duality/Dualities and their ranks \\
	\hline
	Bulk &$Y \sim Y^*$ \quad\quad\quad\quad\quad  $q+1 \sim p+2-q$ \\
	\hline
	Brane &$Y \sim \overline{\underline{Y}}^*+\overline{X}^*+\underline{X}^*$ \quad\quad $q+1 \sim p-q$ \\
	  &$\underline{Y}+X \sim \overline{Y}^*+X^*$ \quad\quad  $q \sim p+1-q$\\
	   &$\overline{Y}+X \sim \underline{Y}^*+X^*$ \quad\quad  $q \sim p+1-q$\\
	 &$ \overline{\underline{Y}}+\overline{X}+\underline{X} \sim Y^*$ \quad\quad  $q-1 \sim p+2-q$\\
	\hline
\end{tabular} \\

Interestingly, the coupled fields in the brane dualities are exactly the blocks appear in the invariant brane action, thus if we take the four dualities (\ref{Hodge duality1}) and (\ref{Hodge duality2})  into our brane action $S_{q}^{(n)}$(\ref{braneaction}) derived for a $q$-form potential, it will automatically equal its dual counterpart $S_{p+1-q}^{*(n)}$:
\begin{eqnarray}
S_{q}^{(n)}
 &=&\frac{-1}{2(q+1)!}\int d^{p+1}x\sqrt{-\hat{g}}
\left(Y^{(n)}_{\mu_{1}\cdots\mu_{q+1}}\right)^{2}+\nonumber\\
&&\frac{-1}{2q!}\int d^{p+1}x\sqrt{-\hat{g}}
\left[\left(\overline{Y}^{(n)}_{\mu_{1}\cdots\mu_{q}}-\frac{m_{y}^{(n)}}{q+1}X^{(n)}_{\mu_{1}\cdots\mu_{q}}\right)^{2}
+\left(\underline{Y}^{(n)}_{\mu_{1}\cdots\mu_{q}}-\frac{m_{z}^{(n)}}{q+1}X^{(n)}_{\mu_{1}\cdots\mu_{q}}\right)^{2}\right]+\nonumber\\
&&\frac{-1}{2(q-1)!}\int d^{p+1}x\sqrt{-\hat{g}}
\left(\underline{\overline{Y}}^{(n)}_{\mu_{1}\cdots\mu_{q-1}}
+\frac{\eta_{(n)}m_{z}^{(n)}}{q}\overline{X}^{(n)}_{\mu_{1}\cdots\mu_{q-1}}
-\frac{\eta_{(n)}m_{y}^{(n)}}{q}\underline{X}^{(n)}_{\mu_{1}\cdots\mu_{q-1}}
\right)^{2}\nonumber\\
&=&\frac{-1}{2(p-q)!}\int d^{p+1}x\sqrt{-\hat{g}}
\left(\underline{\overline{Y}}^{*(n)}_{\mu_{1}\cdots\mu_{p-q}}
+\frac{\eta_{(n)}^{*}m_{z}^{*(n)}}{p+1-q}\overline{X}^{*(n)}_{\mu_{1}\cdots\mu_{p-q}}
-\frac{\eta_{(n)}^{*}m_{y}^{*(n)}}{p+1-q}\underline{X}^{*(n)}_{\mu_{1}\cdots\mu_{p-q}}
\right)^{2}+\nonumber\\
&&\frac{-1}{2(p+1-q)!}\int d^{p+1}x\sqrt{-\hat{g}}
\left[\left(\overline{Y}^{*(n)}_{\mu_{1}\cdots\mu_{p+1-q}}-\frac{m_{z}^{*(n)}X^{*(n)}_{\mu_{1}\cdots\mu_{p+1-q}}}{p+2-q}\right)^{2}
+\left(\underline{Y}^{*(n)}_{\mu_{1}\cdots\mu_{p+1-q}}-\frac{m_{y}^{*(n)}X^{*(n)}_{\mu_{1}\cdots\mu_{p+1-q}}}{p+2-q}\right)^{2}\right]+\nonumber\\
&&\frac{-1}{2(p+2-q)!}\int d^{p+1}x\sqrt{-\hat{g}}
\left(Y^{*(n)}_{\mu_{1}\cdots\mu_{p+2-q}}\right)^{2}
\quad\quad=\quad S_{p+1-q}^{*(n)}.
\end{eqnarray}
In this sense one may conclude that equivalent bulk fields yield equivalent brane fields.

An equally important requirement is that the duality should be compatible with localizability: since according to (\ref{Hodge duality1}) and (\ref{Hodge duality2}), the extra-dimensional parts of the four modes  are (reversely) proportional to their duals', i.e,  $W_i$ to $W^{*}_{4-i}$, two corresponding modes are simultaneously localizable or not, therefore the contradiction mentioned in  this section does not appear.

\section{Conclusion and discussion}\label{conclusion}

In this work, via the gauge-free localization mechanism, we investigated the localization of a $q$-form field in a bulk on a codimension-two $p$-brane.  There turns out to be four types of KK modes: one $q$-form that appears in the usual localization, and in addition, two ($q-1$)-forms, and one ($q-2$)-form. Each type of the modes was found to satisfy two Schr\"{o}dinger equations due to the two codimensions,  so in total we have eight equations for all modes. From these equations and the brane action we observed how the  KK modes obtain masses from the codimensions: the highest mode is free of codimensional index therefore obtains masses form both codimensions; the intermediate mode with index $y$ gains mass from $z$-dimension and the one with index $z$ gains mass from $y$-dimension; the lowest mode has index $xy$ and so has no mass. The mass spectra for modes of different ranks are related by common parameters $\eta_{(n)}$s.

Then we found the effective action on the brane is gauge invariant. Through a multiple of downward couplings, the mass terms in the original action are compensated by one-rank-lower forms, and the action can be rewritten as a sum of such compensated squares. Thus under certain gauge transformations, the brane action is invariant.

Hodge duality is preserved as well. We found the Hodge duality in the bulk naturally reduces to four coupled dualities on the brane, where the coupled brane fields in the brane action match with their duals reversely by order, which makes effective brane actions of the bulk dual fields equivalent.  Incidentally, the codimensional functions of four modes and their dual counterparts' are consistent with each other, so that there is no localizability contradiction. It should be noted that the duality transformation possesses a generality regardless of the mass modes, dualities of zero modes or various massive modes are special aspects of one single entirety.

Furthermore, from the derivation one can realize that, there is nothing special about codimension two, generalizing the results to higher $n$ dimensional reduction is just routine. For example, there will be in total $2^{n}$ types of KK modes, ranking form $q$ to $q-n$, and in the brane action each of them obtains masses form the codimensions whose index it does not contain: for example, the highest-rank KK mode has $n$ mass terms while the lowest dose not have mass. These KK modes couple rank by rank to guarantee action gauge invariance; Hodge duality is preserved in the same manner, by reverse-rank pairing. Even in the situation where the $q$-form fields do not have enough rank for the codimensions or have too much rank for the brane, i.e., $q<n$ or $q>p+1$ , everthing still works well except that there will be fewer KK modes.

Because of emergence of those lower rank KK types, which are typical of the new mechanism,  it is even harder to find solutions for the system, hence to determine localizability. Nevertheless, being free of physical restriction, this general KK decomposition enjoys a primitive mathematical nature, as a result the gauge ivariance and Hodge duality are inheritable through the dimensional reduction. One may understand this from another point of view \cite{Fu2016} that choosing a gauge before the KK decomposition will implicitly eliminate parts of the localization information, so in order to see the whole view, it is reasonable to consider a gauge-free decomposition.

\section*{Acknowledgement}
We sincerely thank Prof. Yu-Xiao Liu for helpful discussions. This work was supported by the Fundamental
Research Funds for the Central Universities (No. Lzujbky-2019-ct06 and No. xzy012019061) and  the National Natural Science Foundation of China (Grants No. 11405121, No. 11705070).

\end{document}